\documentclass[twocolumn,showpacs,preprintnumbers,amsmath,amssymb]{revtex4-1}

\usepackage{graphicx}

\begin{document}

\title{
Condition for directly testing scalar modes of gravitational waves by four detectors
} 
\author{Yuki Hagihara}
\author{Naoya Era}
\author{Daisuke Iikawa}
\author{Naohiro Takeda}
\author{Hideki Asada} 
\affiliation{
${}^{ }$
Graduate School of Science and Technology, Hirosaki University,
Aomori 036-8561, Japan
}

\date{\today}

\begin{abstract} 
General metric theories in a four-dimensional spacetime 
allow at most six polarization states 
(two spin-0, two spin-1 and two spin-2) 
of gravitational waves (GWs). 
If a sky location of a GW source with the 
electromagnetic 
counterpart satisfies 
a single equation that we propose in this paper, both the spin-1 modes 
and spin-2 ones can be eliminated from a certain combination 
of strain outputs at four ground-based GW interferometers  
(e.g. a network of aLIGO-Hanford, aLIGO-Livingston, Virgo and KAGRA), 
where this equation describes curves on the celestial sphere. 
This means that, if a GW source is found in the curve (or its neighborhood practically), 
a direct test of scalar (spin-0) modes separately 
from the other (vector and tensor) modes becomes possible in principle. 
The possibility of such a direct test is thus higher than 
an earlier expectation (Hagihara et al. PRD, 100, 064010, 2019), 
in which they argued that the vector modes could not be completely eliminated. 
We discuss also that adding the planned LIGO-India detector as a fifth detector 
will increase the feasibility of scalar polarization tests. 
\end{abstract}


\maketitle

\section{Introduction}
The greatest achievement in Einstein's theory of general relativity (GR) is 
that our spacetime is not a fixed flat background but becomes a dynamical system, 
and it is described by using a metric in pseudo Riemannian geometry 
\cite{Einstein1916, Einstein1918}. 
GR may be conflict with suggestions from quantum physics and 
string theoretical viewpoints, 
though GR has passed the parameterized post-Newtonian (PPN) tests over a century 
and it is consistent also with aLIGO and Virgo observations of gravitational waves (GWs) 
\cite{Will}. 
The PPN tests are limited within a weak field such as the solar system 
(or mildly relativistic system such as a binary pulsar). 
In this sense, GW observations in a strong field must be important 
for probing new physics beyond GR. 
General metric theories in a four-dimensional spacetime 
allow at most six GW polarization states 
(two spin-0, two spin-1 and two spin-2) \cite{Eardley}. 
Note that two scalar modes (called Breathing and Longitude modes) 
are degenerate for interferometers \cite{Nishizawa2009}.  
Hence, we consider a combination of the two scalar modes 
in this paper.

The first test on the GW polarizations was done for GW150914 
\cite{LIGO2016}. 
This test is inconclusive, because the number of GR polarizations in GR 
was equal to the number of aLIGO detectors. 
The addition of Virgo to the GW detector network 
allowed for the first informative test of GW polarizations for GW170814. 
Their analysis shows that the GW data are described better 
by the pure tensor modes than pure scalar or pure vector modes, 
with Bayes factors in favor of tensor modes of more than 100 and 200, respectively 
 \cite{LIGO2017}. 
A range of tests of GR for GW170817, the first observation of GWs from 
a binary neutron star inspiral, were done by aLIGO and Virgo \cite{LIGO2019}. 
The tests include a test similar to Ref. \cite{LIGO2017} by performing 
a Bayesian analysis of the signal properties with the three detector outputs, 
using the tensor, the vector or the scalar response functions, 
though the signal-to-noise ratio in Virgo was significantly lower than those 
in the two aLIGO detectors. 
Note that the data stream in Virgo still carries information about the signal. 
The prospects for polarization tests were discussed (e.g. \cite{Hayama, Isi2015,Isi2017,Takeda}). 

KAGRA is expected to soon add to the network of GW detectors \cite{LVK}. 
The four noncoaligned GW detectors will allow for better tests of extra GW polarizations 
and stronger constraints on them. 
Generally speaking, the number of the detectors including KAGRA 
is still smaller than the maximum number of the possible polarizations. 
Hagihara et al. found that there exist particular sky positions 
that allow a test of vector modes separately from the other modes, 
because the contributions of possible scalar modes from the GW source 
in a particular sky direction can be canceled out 
in a linear combination of the detectors' outputs \cite{Hagihara2018}. 

Investigating scalar modes is more important than vector modes, 
because many theories of modified gravity, notably scalar-tensor theories,  
have attracted a lot of interest so far \cite{ST-book}. 
Therefore, Hagihara et al. examined whether both vector and tensor modes 
can be perfectly killed in a sky position \cite{Hagihara2019}.  
They did not find such particular sky positions. 
However, there exist some source regions in which the contributions from 
vector modes are not zero but significantly small with killing tensor modes.

The main purpose of the present paper is 
to examine whether only the scalar modes can be extracted 
from a linear combination of the outputs of four detectors. 
We show that, if a GW source is found in a particular sky region, 
the scalar modes can be tested separately from the other (vector and tensor) modes 
in principle. 

This paper is organized as follows. 
In Section II, we discuss how to find a particular linear combination 
of the detector outputs for perfectly killing both the vector mode and the tensor one. 
Section III mentions the arrival time difference between detectors. 
Section IV is devoted to Conclusion. 

Throughout this paper, 
$c$ denotes the speed of light. 
Latin indices $a, b, \cdots$ run from 1 to 4 
corresponding to four detectors. 
We use the Einstein's summation convention 
($A^a B_a = A^1 B_1 + A^2 B_2 + A^3 B_3 + A^4 B_4$). 
$I, J, \cdots$ mean GW polarizations.

\section{Extracting only the scalar modes} 
\subsection{Basic formulation}
Let us imagine four noncoaligned detectors ($a = 1, 2, 3, 4$). 
As is the case of GW events with an 
electromagnetic (EM)  
counterpart such as GW170817 
\cite{GW170817}, 
we assume also that, for a given GW source, we know its sky position.  
By this second assumption, 
we know exactly how to shift the arrival time of the GW 
from detector to detector. 

A general metric theory in a four-dimensional spacetime allows 
at most six polarizations \cite{Eardley}; 
$h_S$ for a spin-0 breathing mode, 
$h_L$ for a spin-0 longitudinal mode, 
$h_V$ and $h_W$ for two spin-1 modes, 
$h_+$ for a spin-2 plus mode 
and 
$h_{\times}$ for a spin-2 cross mode. 
The antenna pattern function of each detector to these polarization modes 
is denoted as 
$F_a^{I}$, where $I = S, L, V, W, +, \times$ 
\cite{PW,footnote1,ST}.  
$F_a^{I}$ is a function of the GW source direction 
and the polarization angle. 
The subscript $a$ of $F_a^I$ for a polarization state 
means a label in the configuration space of the four detectors 
but not in the physical spacetime.  

The strain output at each detector is a superposition as \cite{Nishizawa2009, PW}
\begin{align}
S_a =& F_a^S (h_S - h_L) + F_a^V h_V + F_a^W h_W 
+ F_a^+ h_+ + F_a^{\times} h_{\times} 
\nonumber\\
&+ n_a
\nonumber\\
=& \sum_{I = S}^\times 
F_a^I h_I + n_a ,  
\label{Sa}
\end{align}
where $n_a$ denotes a noise. 

First, we study how to eliminate three polarization modes from 
the signal output in Eq. (\ref{Sa}). 
We introduce the Levi-Civita symbol in the detector configuration space 
$\varepsilon^{abcd}$ 
($a, b, c, d$ take from 1 to 4), where 
$\varepsilon^{1234} = 1$,  $\varepsilon^{abcd}$ is completely antisymmetric, 
and 
the superscripts such as $abcd$ are denoting GW detectors. 
Therefore, $\varepsilon^{abcd}$ is independent of coordinate transformations 
and hence it is a scalar in the physical spacetime. 

By noting the complete antisymmetry of $\varepsilon^{abcd}$, 
one can show 
$(\varepsilon^{abcd} F_a^W F_b^+ F_c^{\times}) F_d^W 
= (\varepsilon^{abcd} F_a^W F_b^+ F_c^{\times}) F_d^+ 
= (\varepsilon^{abcd} F_a^W F_b^+ F_c^{\times}) F_d^{\times} 
=0$. 
Namely, $\varepsilon^{abcd} F_b^W F_c^+ F_d^{\times}$ is 
normal to every of $F_a^W$, $F_a^+$ and $F_a^{\times}$.

We define a projection operator in a space of the antenna pattern functions. 
For $F_a^I$, $F_b^J$ and $F_c^K$, we define 
\begin{equation}
\Pi^{aIJK} \equiv \varepsilon^{abcd} F_b^I F_c^J F_d^K . 
\label{Pi}
\end{equation}
By using this projection operator, we eliminate 
three polarizations from the signal output. 
For $I=W$, $J=+$ and $K=\times$ for example, 
the projection operator becomes 
$\Pi^{aW+\times}$ . 
For this example, $h_W$, $h_+$ and $h_{\times}$ in the strain outputs $\{S_a\}$ are eliminated as 
\begin{align}
\Pi^{aW+\times} S_a 
= 
&\left(\varepsilon^{abcd}F_a^SF_b^WF_c^+F_d^{\times}\right) (h_S-h_L)
\nonumber\\
&+ 
\left(\varepsilon^{abcd}F_a^VF_b^WF_c^+F_d^{\times}\right) h_V 
+ \Pi^{aW+\times} n_a . 
\label{4null-wpc}
\end{align}
We refer to Eq. (\ref{4null-wpc}) as a $W+\times$ null stream. 
By the same way, we can define ten null streams 
for the four detectors as 
$\Pi^{aSVW} S_a, \cdots, \Pi^{aW+\times} S_a$. 
This type of null streams including Eq. (\ref{4null-wpc}) are discussed 
by Chatziioannou et al. \cite{CYC}. 

If the coefficient of $h_V$ in Eq. (\ref{4null-wpc}) 
vanishes in a certain sky region, 
there remains only the spin-0 part in the null steam.  
Thereby, the spin-0 polarization test is possible, 
if a GW source is found in this sky region. 
The vanishing coefficient condition is 
\begin{equation}
\varepsilon^{abcd}F_a^VF_b^WF_c^+F_d^{\times} = 0 . 
\label{epsilon-vwcp}
\end{equation}
This is rewritten in the form of the determinant of a $4 \times 4$ matrix as 
\begin{align}
D_4 
&\equiv
\left|
\begin{array}{cccc}
F_1^V & F_1^W & F_1^+ & F_1^{\times} \\
F_2^V & F_2^W & F_2^+ & F_2^{\times} \\
F_3^V & F_3^W & F_3^+ & F_3^{\times} \\
F_4^V & F_4^W & F_4^+ & F_4^{\times} \\
\end{array}
\right|
\nonumber\\
&= 0 . 
\label{D4}
\end{align}

The components of $h_V$, $h_W$, $h_+$ and $h_{\times}$ are 
dependent on a choice of a reference axis in the transverse plane, 
corresponding to a degree of freedom for a rotation around the GW propagation axis. 
Therefore, one may ask whether the above condition by Eq. (\ref{epsilon-vwcp}) 
is invariant under the rotational transformation. 

We study a rotational transformation of the GW components \cite{PW}, 
where the rotation is considered around the GW propagation axis 
with the rotation angle denoted as $\eta$. 
$h_+$ and $h_{\times}$ are spin 2. 
They are transformed as 
\begin{align}
\left(
\begin{array}{c}
\bar{h}_{+} \\
\bar{h}_{\times}
\end{array}
\right) 
= 
\left(
\begin{array}{cc}
\cos2\eta & \sin2\eta \\
-\sin2\eta & \cos2\eta \\
\end{array}
\right)
\left(
\begin{array}{c}
h_{+} \\
h_{\times}
\end{array}
\right) . 
\label{rotation-hTT}
\end{align}
The bar denotes a quantity after the rotational transformation. 

The antenna pattern functions of each detector 
for $h_+$ and $h_{\times}$ are transformed as 
\begin{align}
\left(
\begin{array}{c}
\bar{F}_a^{+ } \\
\bar{F}_a^{\times}
\end{array}
\right) 
= 
\left(
\begin{array}{cc}
\cos2\eta & -\sin2\eta \\
\sin2\eta & \cos2\eta \\
\end{array}
\right)
\left(
\begin{array}{c}
F_a^{+} \\
F_a^{\times}
\end{array}
\right) . 
\label{rotation-FTT}
\end{align} 

Next, we consider spin 1. 
$h_V$ and $h_W$ are transformed as 
\begin{align}
\left(
\begin{array}{c}
\bar{h}_{V}\\
\bar{h}_{W}
\end{array}
\right) 
= 
\left(
\begin{array}{cc}
\cos\eta & \sin\eta \\
-\sin\eta & \cos\eta \\
\end{array}
\right)
\left(
\begin{array}{c}
h_{V} \\
h_{W}
\end{array}
\right) .
\label{rotation-hVW}
\end{align}
The antenna pattern functions of each detector 
for $h_V$ and $h_W$ are transformed as 
\begin{align}
\left(
\begin{array}{c}
\bar{F}_a^{V} \\
\bar{F}_a^{W}
\end{array}
\right) 
= 
\left(
\begin{array}{cc}
\cos\eta & -\sin\eta \\
\sin\eta & \cos\eta \\
\end{array}
\right)
\left(
\begin{array}{c}
F_a^{V} \\
F_a^{W}
\end{array}
\right) . 
\label{rotation-FVW} 
\end{align} 

By using Eq. (\ref{rotation-FTT}), one can show 
\begin{align}
\varepsilon^{abcd} \bar{F}_c^{+} \bar{F}_d^{\times} 
=& \varepsilon^{abcd} 
(F_c^+ \cos2\eta - F_c^{\times} \sin2\eta) 
\nonumber\\
&
\times (F_d^+ \sin2\eta + F_d^{\times} \cos2\eta) 
\nonumber\\ 
=& 
\varepsilon^{abcd} F_c^{+} F_d^{\times} , 
\label{epsilonFTT}
\end{align}
where $\varepsilon^{abcd}$ is a constant for the rotation. 
Therefore, 
$\varepsilon^{abcd} F_c^{+} F_d^{\times}$ 
is invariant for the rotational transformation. 
By using Eq. (\ref{rotation-FVW}) in the similar manner, we find  
\begin{align}
\varepsilon^{abcd} \bar{F}_c^{V} \bar{F}_d^{W} 
=& 
\varepsilon^{abcd} 
(F_c^V \cos\eta - F_c^W \sin\eta)
\nonumber\\
&
\times (F_d^V \sin\eta + F_d^W \cos\eta) 
\nonumber\\ 
=& 
\varepsilon^{abcd} F_c^V F_d^W . 
\label{epsilonFVW}
\end{align}
Therefore, 
$\varepsilon^{abcd} F_c^V F_d^W$ 
is invariant for the rotational transformation. 

By combining Eqs. (\ref{epsilonFTT}) and (\ref{epsilonFVW}), 
one can show 
\begin{equation}
\varepsilon^{abcd} \bar{F}_a^{V} \bar{F}_b^{W} \bar{F}_c^{+} \bar{F}_d^{\times} 
= 
\varepsilon^{abcd} F_a^{V} F_b^{W} F_c^{+} F_d^{\times} 
\label{epsilonFFFF}
\end{equation}
Therefore,  $D_4$ in Eq. (\ref{D4}) is invariant 
for the rotation around the GW propagation axis.

Eq. (\ref{epsilon-vwcp}), which is equivalent to Eq. (\ref{D4}), describes  
particular sky positions, in which every of the spin-1 ($h^V$ and $h^W$) 
and spin-2 ($h^+$ and $h^{\times}$) parts are eliminated 
in the null stream. 
Namely, Eq. (\ref{4null-wpc}) for such a particular source location becomes
\begin{equation}
\Pi^{aW+\times} S_a 
= \left(\varepsilon^{abcd}F_a^SF_b^WF_c^+F_d^{\times}\right) (h_S-h_L) 
+ \Pi^{aW+\times} n_a . 
\label{4null-wpc-2}
\end{equation}

\begin{figure}
\includegraphics[width=8.6cm]{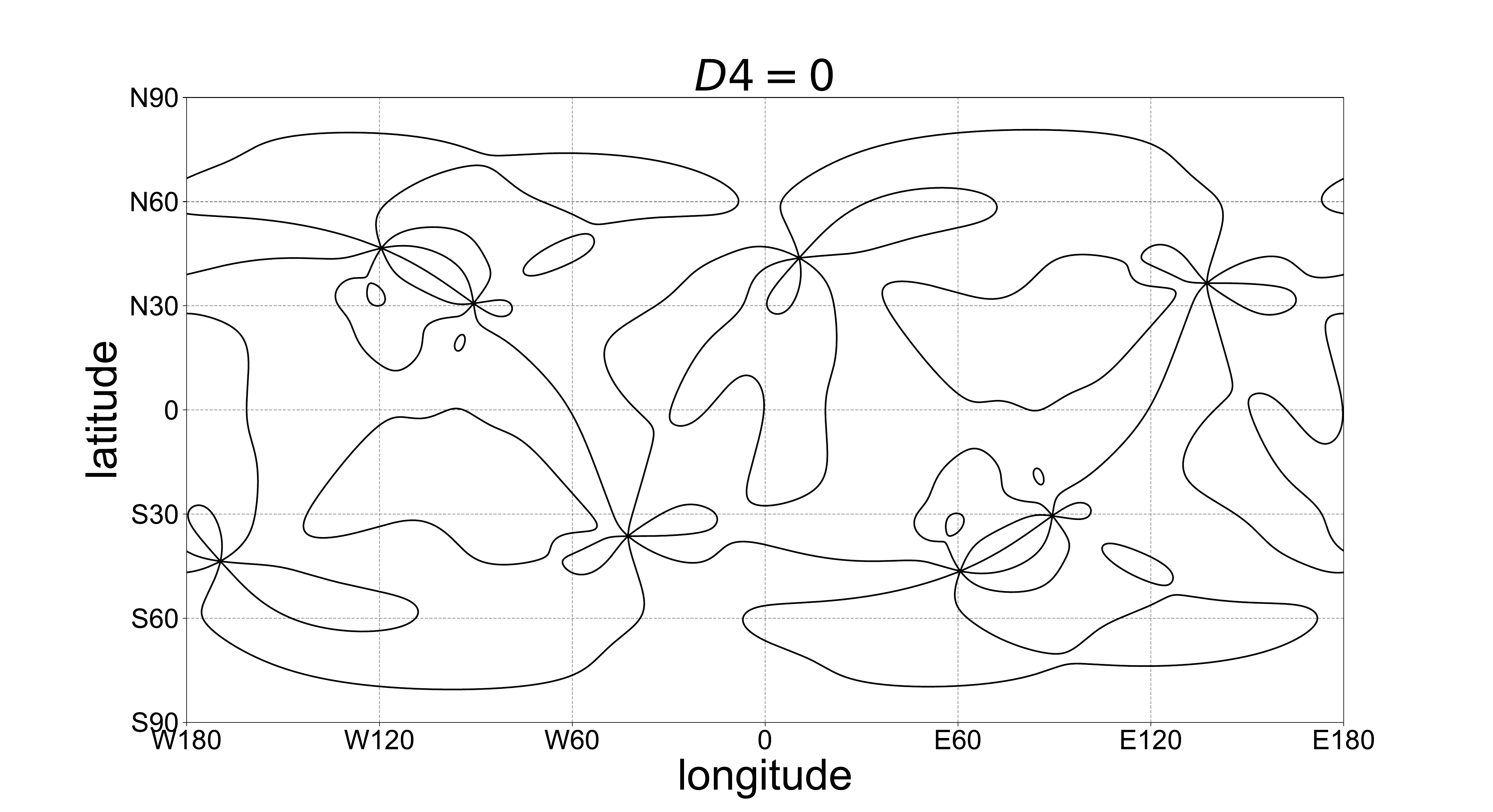}
\caption{
Sky locations for $D_4 = 0$. 
We assume aLIGO, Virgo and KAGRA. 
The vertical axis denotes the latitude and 
the horizontal axis denotes the longitude, 
where the coordinate system is earth-centered. 
Note that this plot does not depend on choices of a polarization angle.
}
\label{figure-D4}
\end{figure}

\begin{figure}
\includegraphics[width=8.6cm]{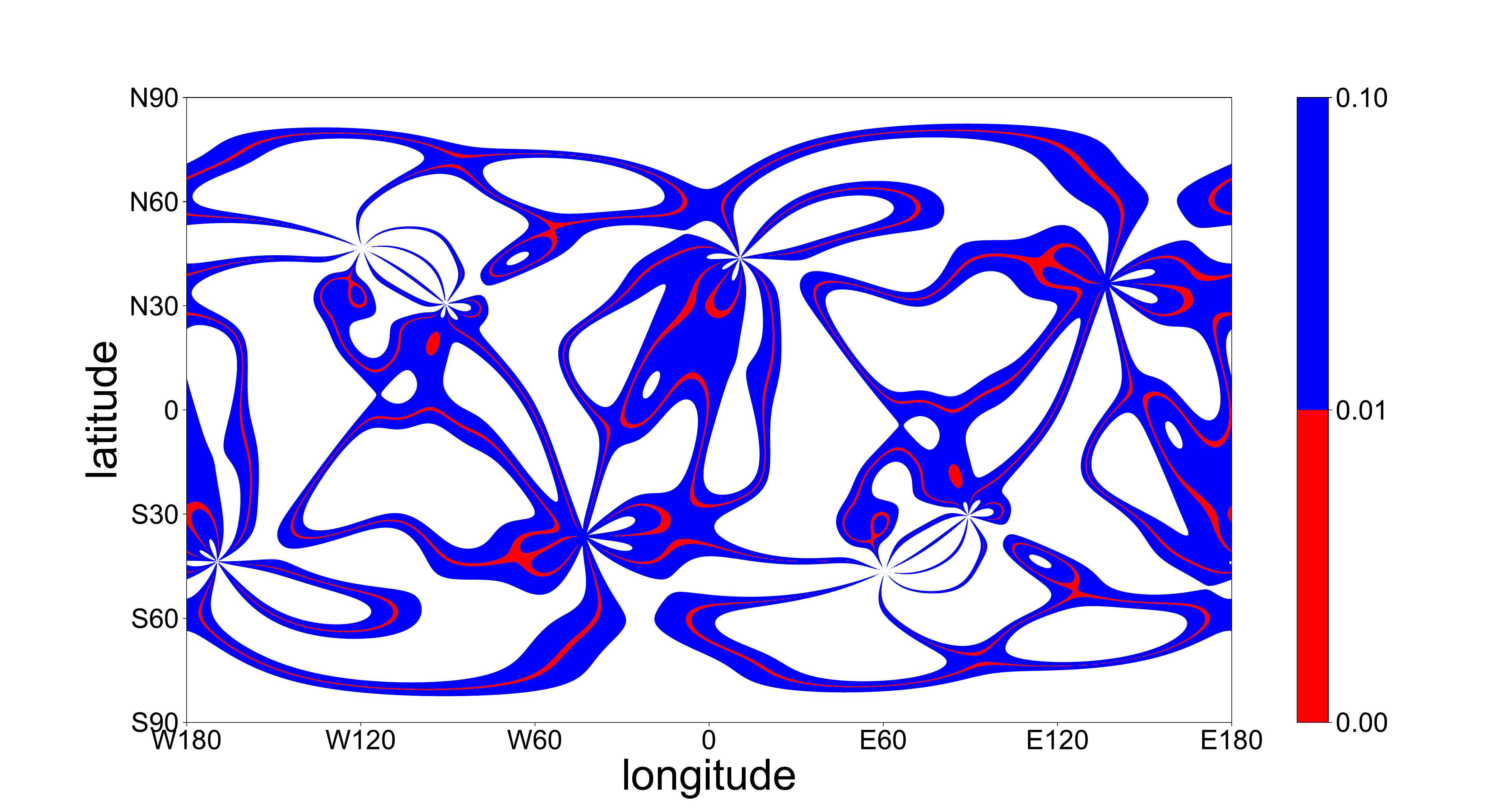}
\caption{
Contour map of $D_4$ 
corresponding to Figure \ref{figure-D4}. 
The red and blue (in color) regions denote $0 \leq 
 |D_4| 
< 0.01$ 
and $0.01 \leq 
 |D_4|  
< 0.1$, respectively. 
Roughly speaking, $|D_4|$ is likely to be $\sim O(1)$, 
because $|F_a^I| \sim O(1)$.  
The area for significantly small $D_4$ such as $0 \leq 
|D_4| 
< 0.01$ 
(red in color) is 
slight but not negligible in this figure. 
}
\label{contour-D4}
\end{figure}

A direct test of the scalar modes becomes possible, 
if the GW source position satisfies Eq. (\ref{D4}). 
This is a main result of this paper. 
See Figure \ref{figure-D4} for sky locations of $D_4 = 0$. 
See also Figure \ref{contour-D4} for a contour map of $D_4$ in the sky.

The fraction of sky area for $0.01 \leq |D_4| < 0.1$ 
(corresponding to the blue region of Figure \ref{contour-D4}) 
is 0.37, 
which means that 
the probability of a GW event 
in a finite range $0.01 \leq |D_4| < 0.1$ 
is 37 percents. 
But a test of scalar polarizations for this case is very weak. 
For a stronger test, $|D_4|$ must be smaller. 
The fraction of sky area for $|D_4| < 0.01$ 
(covered in the red region of Figure \ref{contour-D4}) 
is 0.04. 
Namely, only four percents of GW170817-like events 
satisfy this finite range as $|D_4| < 0.01$, 
which will allow for a stronger direct test of scalar polarizations.

\subsection{Comparison with a {\it three-detector} null stream approach} 
For a comparison, we follow References \cite{Hagihara2018,Hagihara2019} 
to prepare two sets of detectors for four detectors; 
the set (1) is the detectors $a=1, 2, 3$ and  
the other set (2) is $a=2, 3, 4$. 
We define three-dimensional vectors from antenna pattern functions as 
\begin{align}
\vec{F}_{(1)}^{I} &\equiv (F_1^I, F_2^I, F_3^I) , 
\\
\vec{F}_{(2)}^{I} &\equiv (F_2^I, F_3^I, F_4^I) .
\end{align}
We define also vectors for strain outputs and noises as 
\begin{align}
\vec{S}_{(1)} &\equiv (S_1, S_2, S_3) , 
\\
\vec{S}_{(2)} &\equiv (S_2, S_3, S_4) . 
\end{align}
\begin{align}
\vec{n}_{(1)} &\equiv (n_1, n_2, n_3) , 
\\
\vec{n}_{(2)} &\equiv (n_2, n_3, n_4) . 
\end{align}

The outer product as $\vec{F}_{(1)}^{+} \times \vec{F}_{(1)}^{\times}$ 
is perpendicular to both $\vec{F}_{(1)}^{+}$ and $\vec{F}_{(1)}^{\times}$, 
where the outer product is defined in a detector space. 
Therefore, we use it to eliminate the spin-2 $+$ and $\times$ modes 
from the strain outputs. 
\begin{align}
(\vec{F}_{(1)}^{+} \times \vec{F}_{(1)}^{\times}) \cdot 
\vec{S}_{(1)} 
=& 
[(\vec{F}_{(1)}^{+} \times \vec{F}_{(1)}^{\times}) \cdot \vec{F}_{(1)}^S] (h_S - h_L) 
\nonumber\\
&+ [(\vec{F}_{(1)}^{+} \times \vec{F}_{(1)}^{\times}) \cdot \vec{F}_{(1)}^V] h_V 
\nonumber\\
&+ [(\vec{F}_{(1)}^{+} \times \vec{F}_{(1)}^{\times}) \cdot \vec{F}_{(1)}^W] h_W 
\nonumber\\
&+ (\vec{F}_{(1)}^{+} \times \vec{F}_{(1)}^{\times}) \cdot 
\vec{n}_{(1)} , 
\label{S1}
\\ 
(\vec{F}_{(2)}^{+} \times \vec{F}_{(2)}^{\times}) \cdot 
\vec{S}_{(2)} 
=&  
[(\vec{F}_{(2)}^{+} \times \vec{F}_{(2)}^{\times}) \cdot \vec{F}_{(2)}^S] (h_S - h_L) 
\nonumber\\
&+ [(\vec{F}_{(2)}^{+} \times \vec{F}_{(2)}^{\times}) \cdot \vec{F}_{(2)}^V] h_V 
\nonumber\\
&+ [(\vec{F}_{(2)}^{+} \times \vec{F}_{(2)}^{\times}) \cdot \vec{F}_{(2)}^W] h_W 
\nonumber\\
&+ (\vec{F}_{(2)}^{+} \times \vec{F}_{(2)}^{\times}) \cdot 
\vec{n}_{(2)} , 
\label{S2}
\end{align}
where $\cdot$ denotes the inner product. 
These equations are often called null streams in the literature 
\cite{GT,WS,CYC,null-papers}. 
We refer to them as {\it tensor} null streams, 
because only the spin-2 modes are completely killed. 
Tensor null streams are originally for three detectors \cite{ST}.

\begin{figure}
\includegraphics[width=8.6cm]{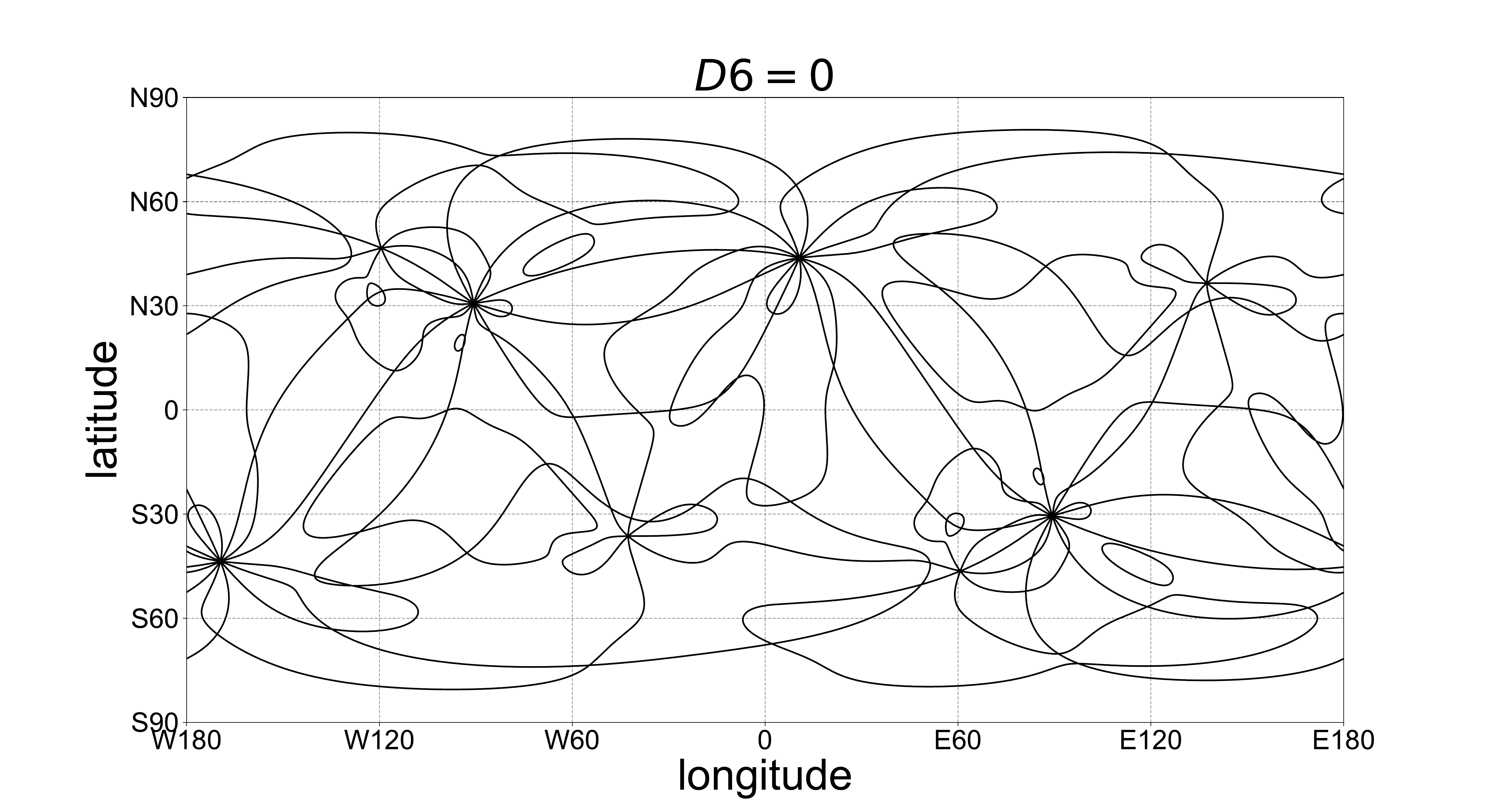}
\caption{
Plot for $D_6 = 0$, corresponding to Figure \ref{figure-D4}. 
The vertical axis denotes the latitude and 
the horizontal axis denotes the longitude. 
}
\label{figure-D6}
\end{figure}

If and only if the antenna pattern functions satisfy 
\begin{align}
D_6^{+\times}  
\equiv& 
[(\vec{F}_{(1)}^{+} \times \vec{F}_{(1)}^{\times}) \cdot \vec{F}_{(1)}^V]
[(\vec{F}_{(2)}^{+} \times \vec{F}_{(2)}^{\times}) \cdot \vec{F}_{(2)}^W]
\nonumber\\
&
- [(\vec{F}_{(1)}^{+} \times \vec{F}_{(1)}^{\times}) \cdot \vec{F}_{(1)}^W] 
[(\vec{F}_{(2)}^{+} \times \vec{F}_{(2)}^{\times}) \cdot \vec{F}_{(2)}^V] 
\nonumber\\
=& 0 , 
\label{D6}
\end{align}
$[(\vec{F}_{(1)}^{+} \times \vec{F}_{(1)}^{\times}) \cdot \vec{F}_{(1)}^V] h_V 
+ [(\vec{F}_{(1)}^{+} \times \vec{F}_{(1)}^{\times}) \cdot \vec{F}_{(1)}^W] h_W$ 
in the right-hand side of Eq. (\ref{S1}) 
is always proportional to 
$[(\vec{F}_{(2)}^{+} \times \vec{F}_{(2)}^{\times}) \cdot \vec{F}_{(2)}^V] h_V 
+ [(\vec{F}_{(2)}^{+} \times \vec{F}_{(2)}^{\times}) \cdot \vec{F}_{(2)}^W] h_W$ 
in the right-hand side of Eq. (\ref{S2}) 
for any $h_V$ and $h_W$. 
Therefore, there exists a linear combination of 
Eqs. (\ref{S1}) and (\ref{S2}), 
such that the spin-1 polarizations also can be eliminated. 
The resultant combination contains only the spin-0 mode with 
eliminating the other (spin-1 and spin-2) modes. 
It seems that this is the same as Eq. (\ref{4null-wpc-2}) 
which was derived from Eq. (\ref{4null-wpc}). 

Is Eq. (\ref{D6}) equivalent to Eq. (\ref{epsilon-vwcp})? 
No. 
This is because 
$D_6^{+\times}$ in Eq. (\ref{D6}) is of sixth degree in antenna pattern functions, 
while $D_4$ in Eq. (\ref{D4}) is of fourth degree. 
$D_6^{+\times}$ is factorized as 
\begin{equation}
D_6^{+\times} = D_4 D_2^{+\times} , 
\label{D6D4}
\end{equation}
where we define 
\begin{equation}
D_2^{+\times} 
\equiv 
\left|
\begin{array}{cc}
F_2^+ & F_2^{\times} \\
F_3^+ & F_3^{\times} \\
\end{array}
\right| . 
\end{equation}

If 
\begin{equation}
D_2^{+\times} = 0 ,
\label{D2}
\end{equation}
$D_6$ vanishes even if $D_4 \neq 0$. 
$D_2^{+\times} = 0$ is a case that tensor antenna pattern functions 
of the second and third detectors are degenerate. 
See Figure \ref{figure-D6} for sky locations of $D_6 = 0$. 
For the case of $D_4 \neq 0$, there remains $h^V$  
in the right-hand side of Eq. (\ref{4null-wpc}). 
This means that the case of $D_2^{+\times} = 0$ does not lead to a null stream. 
Therefore, the present formulation to four detectors 
improves an earlier approach \cite{Hagihara2018, Hagihara2019} 
combining two tensor null streams. 

Hagihara et al. (2019) considered the two null streams 
as Eqs. (\ref{S1}) and (\ref{S2}) \cite{Hagihara2019}. 
They examined whether the four coefficients of $h_V$ or $h_W$ can 
simultaneously vanish. 
However, vanishing of the four coefficients is too strong. 
For a direct test of the scalar modes, 
it is enough that two coefficients vanish.

\subsection{Adding a fifth detector}
Planned LIGO-India is a fifth detector \cite{LIGO-India-HP}. 
One of the largest merits of LIGO-India (labeled as $a=$I) 
comes from its geographic factor, 
namely being very distant from the other detectors HLVK. 
LIGO-H and LIGO-L detectors are approximately aligned. 
Therefore, adding LIGO India is expected to help break some degeneracy 
between H and L. 

\begin{figure}
\includegraphics[width=8.6cm]{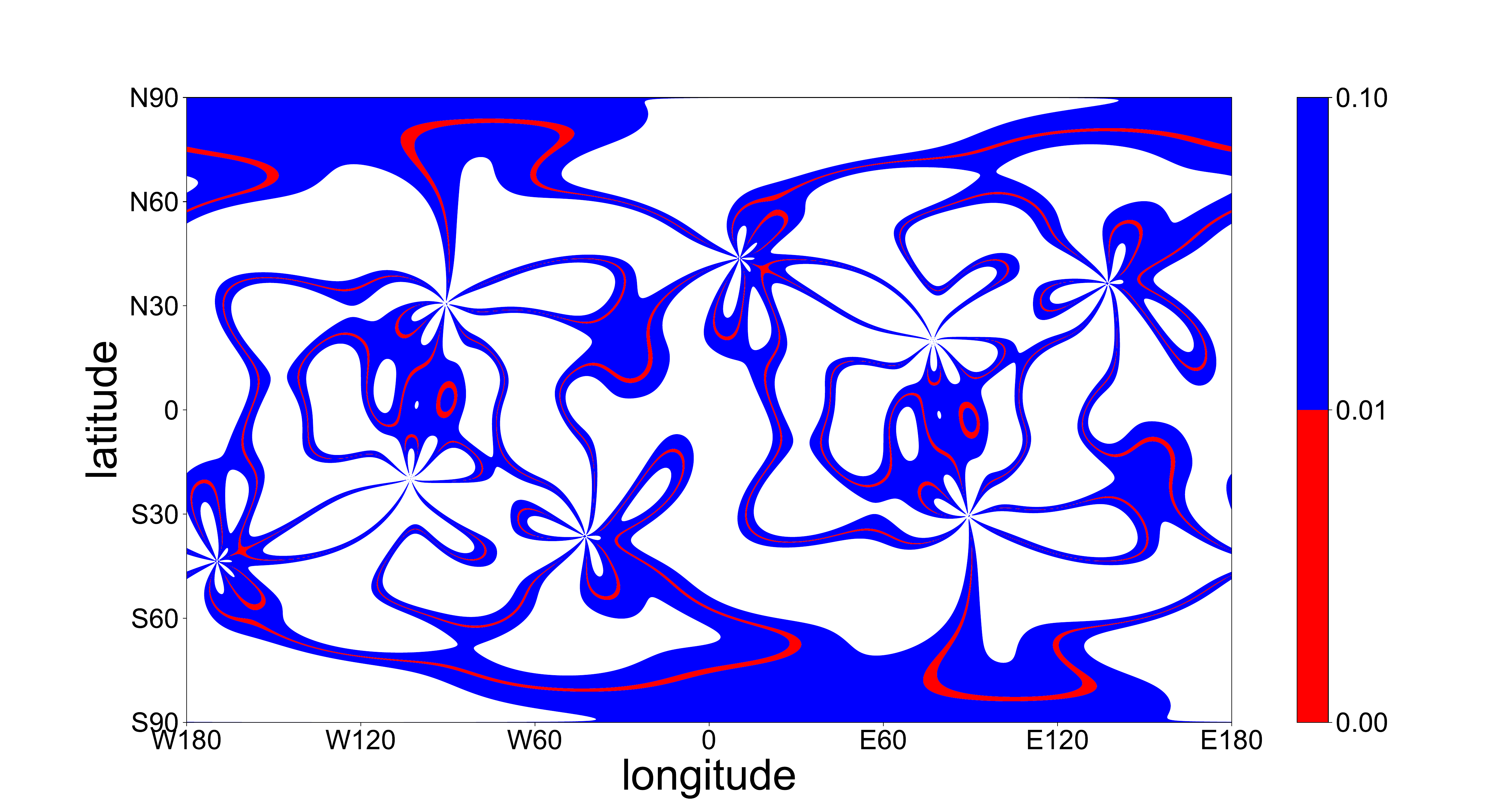}
\caption{
Contour map of $D_4$ for LVKI network, 
corresponding to Figure \ref{contour-D4} for HLVK. 
The coordinates of the LIGO-India detector 
are approximated by those of the Hingoli city 
($19.72^{\circ}$N, $77.15^{\circ}$E) 
\cite{LIGO-India-HP} 
and we assume that 
the detector arms are alined to the east and north directions, respectively,  
for its simplicity. 
The red and blue (in color) regions denote $0 \leq |D_4| < 0.01$ 
and $0.01 \leq |D_4| < 0.1$, respectively. 
}
\label{contour-D4-LVKI}
\end{figure}

By constructing a four-detector null stream including LIGO-India instead of H, 
we compute $D_4$ for LVKI. 
The LIGO-India detector is under planning. 
The detailed information on the detector is not currently open to public. 
Therefore, when computing the antenna pattern function of 
the LIGO-India detector, 
the coordinates of the LIGO-India detector 
are approximated by those of the Hingoli city 
($19.72^{\circ}$N, $77.15^{\circ}$E) 
\cite{LIGO-India-HP} 
and we assume that 
the detector arms are alined to the east and north directions, respectively,  
for its simplicity. 
See Figure \ref{contour-D4-LVKI} for a contour map of $D_4$ for LVKI. 

The areas covered in the red and blue (in color) region 
in Figure \ref{contour-D4-LVKI} are
comparable to 
those in Figure \ref{contour-D4} for HLVK. 
The sky fraction of the red region ($|D_4| < 0.01$) in Figure \ref{contour-D4-LVKI}
is 0.03. 
The fraction of the blue region ($0.01 \leq |D_4| < 0.1$) 
is 0.33. 
In Figure \ref{contour-D4}, they are 0.04 and 0.37, respectively. 
We consider also different sets of detectors to evaluate $D_4$. 
See Table \ref{table-1}. 

\begin{table}
\caption{
Area fraction in the sky for $|D_4| < 0.01$ and $0.01 \leq |D_4| < 0.1$. 
We consider five sets of HLVK, LVKI, HVKI, HLKI and HLVI. 
}
\begin{center}
\begin{tabular}{l|c|c}
 & $|D_4| < 0.01$ & $0.01 \leq |D_4| < 0.1$ \\
\hline
HLVK & 0.04 & 0.37 \\
LVKI & 0.03 & 0.33 \\
HVKI & 0.03 & 0.33 \\
HLKI & 0.05 & 0.43 \\
HLVI & 0.04 & 0.37
\end{tabular}
\end{center}
\label{table-1}
\end{table}

We consider two different arm directions by $30$-degree rotation from the east direction 
and $45$-degree rotation (namely north-east direction).  
See Table \ref{table-2}. 

\begin{table}
\caption{
Area fraction in the sky for $|D_4| < 0.01$ and $0.01 \leq |D_4| < 0.1$. 
We consider three cases of the arm directions for LVKI. 
}
\begin{center}
\begin{tabular}{l|c|c}
 & $|D_4| < 0.01$ & $0.01 \leq |D_4| < 0.1$ \\
 \hline
LVKI $0^{\circ}$ (East) & 0.03 & 0.33 \\
LVKI $30^{\circ}$ & 0.04 & 0.35 \\
LVKI $45^{\circ}$ (North-east) & 0.04 & 0.32
\end{tabular}
\end{center}
\label{table-2}
\end{table}

These calculations show that 
replacing one of HLVK by LIGO-India in a four-detector null stream 
does not so much affect the area fraction. 
Namely, the area fraction is almost independent of a choice of a detector set 
in the four-detector null stream. 
But adding LIGO-India gives us four more sets of four-detector null streams. 
Roughly speaking, therefore, the total area fraction for HLVKI becomes five-times larger 
than only the HLVK network. 

Furthermore, we should stress  
that Eq. (\ref{Sa}) for strain outputs ($a=1, \cdots, 5$) 
from the five detectors including 
LIGO-India can be always solved for the five modes 
$h_S - h_L$, $h_V$, $h_W$, 
$h_+$ and $h_{\times}$. 
This means that, in principle, adding LIGO-India will allow for 
a direct test of each GW polarization for {\it any} sky region. 
This would be an important step in testing our gravitational theories.

\section{On the arrival time of extra polarization modes}
In discussions from Eq. (\ref{rotation-hTT}) to Eq. (\ref{4null-wpc-2}), 
we assume that the speeds of the same spin modes are identical. 
But a different spin mode may travel at different speed. 
We denote the speed of spin-$h$ mode ($h=0, 1, 2$) as $c_h$. 
We examine whether the shift of the arrival time difference from detector to detector 
should be changed for extra polarization modes. 

At the Earth, 
the arrival time difference between the spin-$h$ mode and the light signal is 
\begin{equation}
\delta t_h 
= \frac{D}{c} - \frac{D}{c_h} , 
\label{deltat}
\end{equation}
where $D$ is the distance to the source. 
The arrival time difference between two detectors ($a$ and $b$) 
for the spin-$h$ mode is defined as 
\begin{equation}
\Delta t_{ab} \equiv \frac{D_a - D_b}{c_h} , 
\label{Deltat}
\end{equation}
where $D_a$ and $D_b$ denote the distance from the source 
to the detector $a$ and $b$, respectively. 
By combining Eqs. (\ref{deltat}) and (\ref{Deltat}), 
we eliminate $c_h$ to obtain 
\begin{equation}
\Delta t_{ab} = \frac{D_a - D_b}{c} 
- \frac{D_a - D_b}{D} \delta t_h . 
\label{Deltat2}
\end{equation}

The first term in the right-hand side of Eq. (\ref{Deltat2}) 
is the normal arrival time difference between detectors, 
which has been already taken into account in the GW data analysis. 
The second term is due to the deviation from the light speed. 
The arrival time difference between detectors by $c_h - c$ 
is roughly estimated as 
\begin{align}
\left| \frac{D_a - D_b}{D} \delta t_h \right| 
\sim & 
2 \times 10^{-14} \mbox{sec.}
\nonumber\\
& 
\times 
\left(\frac{|D_a - D_b|}{6 \times 10^3 \mbox{km}}\right)
\left(\frac{40 \mbox{Mpc}}{D}\right)
\left(\frac{|\delta t_h|}{3600 \mbox{sec.}}\right) . 
\end{align}
We assume $|D_a - D_b| \sim $ the Earth size and 
the data analysis duration for one GW event is $\sim 3600$ seconds for the simplicity. 
If the extra mode arrives much later, say a few weeks later, 
we can hardly recognize that it came from the same event. 
A few-hours delay may be identified as the same event in data analysis. 
Therefore, a case that a hypothetical scalar is nearly massless 
can be tested in the present approach, while a scalar with heavy mass 
is beyond the reach of the present method because of large delay. 
If we analyze the data for testing extra modes 
arriving before an hour later, the correction to 
the arrival time difference between detectors for $D \sim 40$Mpc is 
around $\sim 10^{-14}$ seconds, which may be currently negligible 
in the data analysis of waves with the frequency band around a few kHz.

\section{Conclusion} 
We considered strain outputs at four noncoaligned detectors 
such as a network of  
aLIGO-Hanford, aLIGO-Livingston, Virgo and KAGRA \cite{LVK}. 
Generally speaking, five unknowns $\{h_S-h_L, h_V, h_W, h_+, h_{\times}\}$ 
cannot be determined from four outputs $\{S_1, S_2, S_3, S_4\}$. 
If a sky location of a GW source with the EM counterpart satisfies 
a single equation that we proposed in this paper, 
both the spin-1 modes and spin-2 ones 
can be eliminated from a certain combination 
of strain outputs at the four detectors, 
where this equation describes curves on the celestial sphere. 
If a GW source is found in the curve (or its neighborhood practically), 
a direct test of the scalar modes separately 
from the other (vector and tensor) modes becomes possible in principle. 
The possibility of such a direct test is higher than 
the earlier expectation (Hagihara et al. PRD, 100, 064010, 2019), 
which argued that the vector modes could not be completely eliminated.  
We discussed also that 
adding the planned LIGO-India detector as a fifth detector 
would significantly increase the feasibility of scalar polarization tests. 
Detailed numerical simulations with using binary models 
will be left for future work.

\begin{acknowledgments}
We would like to thank Atsushi Nishizawa for useful discussions. 
We would like to thank 
Hideyuki Tagoshi and Jishnu Suresh 
for the information on the current status of the planned LIGO-India detector. 
We wish to thank Seiji Kawamura, Kipp Cannon, Nobuyuki Kanda, 
and Yousuke Itoh for stimulating conversations. 
We thank 
Yuuiti Sendouda and Toshiaki Ono  
for the useful conversations. 
H. A. is supported 
in part by Japan Society for the Promotion of Science (JSPS) 
Grant-in-Aid for Scientific Research, 
No. 17K05431, 
and 
in part by Ministry of Education, Culture, Sports, Science, and Technology,  
No. 17H06359.  
\end{acknowledgments}


\begin{thebibliography}{99}
\bibitem{Einstein1916}
A. Einstein, 
Sitzungsber. Preuss. Akad. Wiss. Berlin (Math. Phys.) 
{\bf 1916}, 688 (1916). 
\bibitem{Einstein1918}
A. Einstein, 
Sitzungsber. Preuss. Akad. Wiss. Berlin (Math. Phys.) 
{\bf 1918}, 154 (1918). 
\bibitem{Will}
C. M. Will, Living Rev. Relativity, 17, 4 (2014). 
\bibitem{Eardley}
D. M. Eardley, D. L. Lee, A. P. Lightman, R. V. Wagoner, and C. M. Will, 
Phys. Rev. Lett. {\bf 30}, 884 (1973).
\bibitem{Nishizawa2009}
A. Nishizawa, A. Taruya, K. Hayama, S. Kawamura, and M. A. Sakagami, 
Phys. Rev. D {\bf 79}, 082002 (2009). 
\bibitem{LIGO2016}
B. P. Abbott et al. (Virgo and LIGO Scientific Collaborations), 
Phys. Rev. Lett. {\bf 116}, 221101 (2016). 
\bibitem{LIGO2017}
B. P. Abbott et al. (Virgo and LIGO Scientific Collaborations), 
Phys. Rev. Lett. {\bf 119}, 141101 (2017). 
\bibitem{LIGO2019}
B. P. Abbott et al. (Virgo and LIGO Scientific Collaborations), 
Phys. Rev. Lett. {\bf 123}, 011102 (2019).  
\bibitem{Hayama}
K. Hayama, and A. Nishizawa, Phys. Rev. D {\bf 87}, 062003 (2013).
\bibitem{Isi2015}
M. Isi, A. J. Weinstein, C. Mead, and M. Pitkin, 
Phys. Rev. D 91, 082002 (2015).
\bibitem{Isi2017}
M. Isi, M. Pitkin, and A. J. Weinstein, 
Phys. Rev. D {\bf 96}, 042001 (2017). 
\bibitem{Takeda}
H. Takeda, A. Nishizawa, Y. Michimura, K. Nagano, K. Komori, 
M. Ando, and K. Hayama, 
Phys. Rev. D {\bf 98}, 022008 (2018).
\bibitem{LVK}
B. P. Abbott, 
Living. Rev. Relativ., {\bf 21}, 3 (2018).
\bibitem{Hagihara2018}
Y. Hagihara, N. Era, D. Iikawa, and H. Asada, 
Phys. Rev. D {\bf 98}, 064035 (2018).  
\bibitem{ST-book}
Y. Fujii, and K. Maeda, 
{\it Scalar-Tensor Theory of Gravitation}, 
(Cambridge Univ. Press, UK. 2008). 
\bibitem{Hagihara2019}
Y. Hagihara, N. Era, D. Iikawa, A. Nishizawa, and H. Asada, 
Phys. Rev. D {\bf 100}, 064010 (2019).  
\bibitem{GW170817}
B. P. Abbott, et al., 
Astrophys. J. Lett. {\bf 848}, L12 (2017);
B. P. Abbott, et al., 
Astrophys. J. Lett. 848, L13 (2017). 
\bibitem{PW}
E. Poisson, and C. M. Will, 
{\it Gravity}, (Cambridge Univ. Press, UK. 2014). 
\bibitem{footnote1}
The present paper follows 
Chapter 13 in Poisson and Will \cite{PW} 
to define the GW antenna patterns.  
See Nishizawa et al. \cite{Nishizawa2009} and 
also a pioneering work for purely TT waves by Schutz and Tinto 
\cite{ST}. 
Note that their definitions are a little different from each other 
\cite{PW, Nishizawa2009}. 
\bibitem{ST}
B. F. Schutz, and M. Tinto, 
Mon. Not. R. Astr. Soc. {\bf 224}, 131 (1987). 
\bibitem{CYC}
K. Chatziioannou, N. Yunes, and N. Cornish, 
Phys. Rev. D {\bf 86}, 022004 (2012). 
\bibitem{GT}
Y. G\"ursel, and M. Tinto, 
Phys. Rev. D {\bf 40}, 3884 (1989). 
\bibitem{WS}
L. Wen, and B. F. Schutz, 
Class. Quant. Grav. {\bf 22}, S1321 (2005). 
\bibitem{null-papers}
S. Chatterji, A. Lazzarini, L. Stein, P. J. Sutton, A. Searle, and M. Tinto,  
Phys. Rev. D {\bf 74}, 082005 (2006). 
\bibitem{LIGO-India-HP}
http://www.ligo-india.in/environmental-clearance-given-to-ligo-india/
\end{thebibliography}
\end{document}